\documentclass[twocolum]{aa}

\newif\ifpdf
\ifx\pdfoutput\undefined
\pdffalse % we are not running PDFLaTeX
\else
\pdfoutput=1 % we are running PDFLaTIn aneX
\pdftrue
\fi

\ifpdf
\usepackage[pdftex]{graphicx}
\else
\usepackage{graphicx}
\fi
\usepackage{txfonts}
\begin{document}
   \title{Bimodal spectral variability of \object{Cygnus~X-1} in an intermediate state}

   \author{J. Malzac
          \inst{1}
          \and
          P.O.~Petrucci\inst{2}
          \and
          E.~Jourdain\inst{1}
           \and
          M.~Cadolle Bel\inst{3,4}
          \and
          P.~Sizun\inst{3}    
          \and
          G.~Pooley\inst{5}
          \and
          C.~Cabanac\inst{2}
          \and
          S.~Chaty\inst{6}
          \and
          T.~Belloni\inst{7}
          \and
          J.~Rodriguez\inst{3,6,8}
          \and
          J.P.~Roques\inst{1}
          \and
           P.~Durouchoux\inst{3}
          \and
          A.~Goldwurm\inst{3,4}
          \and
          P.~Laurent\inst{3,4}
          }

   \offprints{J. Malzac}

   \institute{Centre d'Etude Spatiale des Rayonnements (CNRS/UPS/OMP), F-31028
Toulouse, France
              \email{Julien.Malzac@cesr.fr}
          \and
              Laboratoire d'Astrophysique Observatoire de Grenoble, BP 53
F-38041 GRENOBLE C\'edex 9, France
          \and
              Service d'Astrophysique, CEA-Saclay, Bat. 709,
L'Orme des Merisiers, F-91 191 Gif-sur-Yvette, Cedex, France
          \and
          APC-UMR 7164, 11 place M. Berthelot, 75231 Paris, France
           \and
              Cavendish Laboratory, University of Cambridge, Madingley
Road, Cambridge CB3 0HE, UK
          \and
AIM - Astrophysique Interactions Multi-\'echelles
(Unit\'e Mixte de Recherche 7158 CEA/CNRS/Universit\'e Paris 7 Denis Diderot),
CEA-Saclay, B\^at. 709,
L'Orme des Merisiers, F-91 191 Gif-sur-Yvette Cedex, France
           \and
               INAF - Osservatorio Astronomico di Brera, via
E. Bianchi 46, 23807 Merate, Italy
           \and
             {\it {\it INTEGRAL}} Science Data Center, Chemin d'\'Ecogia 16, 1290 Versoix, Switzerland     
             }

   \date{Received September 15, 1996; accepted March 16, 1997}

   \abstract{
  We report the results of an observation of \object{Cygnus~X-1}
 performed on June 7-11 2003 with  {\it INTEGRAL} 
 { that we combine with simultaneous radio observations with the Ryle telescope.}
 Both spectral and variability properties of
the source indicate that \object{Cygnus~X-1} was in an Intermediate State.
The {\it INTEGRAL} spectrum shows 
a high-energy cut-off or break around 100 keV. The shape of this cut-off 
differs from pure thermal Comptonisation, suggesting the presence of 
a non-thermal component at higher energies.
The average broad band spectrum is well represented
by hybrid thermal/non-thermal Comptonisation models. 
However, models with mono-energetic injection, or models 
with an additional soft component are favoured 
over standard power-law acceleration models.
During the 4 day long observation 
the broad band (3--200 keV) luminosity varied by up to a factor of 2.6 and 
the source showed an important spectral variability. 
A principal component analysis demonstrates that most of this variability
 occurs through 2 independent modes. The first mode consists in 
 changes in the overall luminosity on time scale of hours
 with almost constant spectra (responsible for 68 \% of the variance) 
 { that are strikingly uncorrelated 
 with the variable radio flux.}
 We interpret this variability mode as variations of the dissipation rate in the corona, 
 possibly associated with magnetic flares. 
 The second variability mode consists in a pivoting of the spectrum around
  $\sim$10 keV (27 \% of the variance). It acts on a longer time-scale: 
  initially soft, the spectrum hardens in the first part of the observation and then softens again.
This pivoting pattern is strongly correlated with the radio (15 GHz) emission:
 radio fluxes are stronger when the {\it INTEGRAL} spectrum is harder 
 We propose that the pivoting mode represents
 a 'mini' state transition from a nearly High Soft State to a nearly Low Hard State , and back.
 This mini-transition would be caused by
 changes in the soft cooling photons flux in the hot
 Comptonising plasma associated with an increase of the
  temperature of the accretion disc. 
 { The jet power then appears to be anti-correlated with the disc luminosity
   and unrelated to the coronal power.
  This is in sharp contrast with previous results obtained for the Low Hard State, suggesting
   a different mode of coupling between the jet, the cold disc, 
   and the corona in Intermediate States.}
   From this interpretation we also infer that the bolometric luminosity
  jumps by a factor of about 2 during the transition hard to soft, suggesting a radiatively
   inefficient accretion flow in the Low Hard State.
  
   }

   \maketitle
%
%________________________________________________________________

\section{Introduction}

\begin{figure}[!t]
\centering
\includegraphics[width=\linewidth]{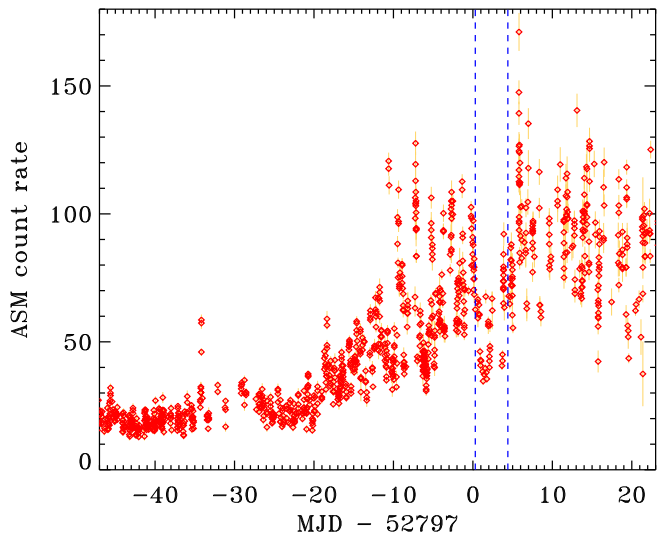}
\caption{{\it RXTE/ASM} light curve of \object{Cygnus~X-1}. The time of the {\it INTEGRAL}
observation  is delimited by the vertical dashed lines \label{fig:asm}}
\end{figure}

\object{Cygnus~X-1} is the prototype of black hole candidates. 
Since its discovery in 1964 (Bowyer et al. 1965), it has been intensively observed by 
all the high-energy instruments, from soft X-rays to $\gamma$-rays. 
It is a persistent source most often observed
 in the so-called Low Hard State (hereafter LHS), characterised 
by a relatively low flux in the soft X-rays ($\sim$ 1 keV) and a
high flux in the hard X-rays ($\sim 100$ keV).
In the LHS, the high-energy spectrum can be roughly described
 by a power-law 
with spectral index $\Gamma$ varying in the range 1.4-2.2, 
and a nearly exponential cut-off at a characteristic energy 
$E_{\rm c}$ of a few hundred keV (see e.g. Gierlinski et al. 1997).
Occasionally, the source switches to the High Soft State (HSS). The
high-energy power-law is then much softer ($\Gamma > 2.4$) and the
bolometric luminosity is dominated by 
a thermal component peaking at a few keV.
Finally, there are also Intermediate States (hereafter IMS)
in  which the source exhibits a
relatively soft hard X-ray spectrum ($\Gamma \sim 2.1-2.3$) and a
moderately strong soft thermal component (Belloni et al. 1996; Mendez \& van der Klis 1997). 
The IMS often,
but not always, appears when the source is about to switch from one
state to the other.
When it is not associated with a
state transition, it is interpreted as a 
`failed state transition'.
Until 1998, the source used to spend nearly 90 \% of its time in the LHS. In the recent years however 
      there have been more IMSs and soft states 
(see Zdziarski et al. 2002,  Pottschmidt et al. 2003, Gleissner et al. 2004).

Simultaneous radio/X-ray and high-energy observations of \object{Cygnus~X-1} and other sources 
have shown that the X-ray LHS is correlated 
with a strong radio emission which is consistent with arising from a jet (Fender 2001).
In contrast, during HSS episodes the source appears to be radio
 weak (Brocksopp et al. 1999).  The presence of a compact jet in the LHS 
 was confirmed by Stirling et al. (2001) who presented evidence 
 for an extended and collimated radio structure on milliarsecond scales.
 
State transitions are generally interpreted as being associated with changes 
 in the geometry of the accretion flow. In the HSS the geometrically thin optically 
 thick disk (Shakura \& Sunyaev 1973) extends down to the last stable orbit. 
 The spectrum is dominated by the thermal disc
 component and peaks at a few keV. The hard X-ray emission is then believed to be produced 
 in a non-thermal corona above and below the disc (Gierli{\'n}ski et al. 1999, hereafter G99). 
  In the LHS the geometrically thin disc is truncated at a few hundred Schwarschild radii, 
  the innermost part of the accretion flow forms a geometrically thick optically thin and
   hot disc (Shapiro et al. 1976; Narayan \& Yi 1994)
    where high-energy radiation is produced trough thermal Comptonisation.
 During  spectral transitions to the HSS  the inner radius of the cold accretion disk decreases.
 This reduction of the inner disc radius is associated with either
  the cold disk penetrating inside 
  the hot inner flow, or the  later collapsing into an optically thick accretion disk 
  with small active regions of hot plasma on top of it 
   (Zdziarski et al. 2002). In both cases the enhanced soft photon flux from the disk 
   tends to cool down the hot phase, leading to softer spectra.
   An alternative possibility would be that in both LHS and HSS, the geometry is that of a
   corona above a cold standard accretion disc (Bisnovatyi-Kogan \& Blinnikov 1976;
    Haard \& Maraschi 1993). In the LHS
   the coronal plasma is essentially thermal and the cold disc faint because most of the power 
   is transported away and dissipated in the corona and a strong outflow 
   (Beloborodov 1999; Malzac, Beloborodov \& Poutanen 2001; Merloni \& Fabian 2002; 
   Ferreira et al. 2005),
   while in the HSS most of the power is dissipated into the cold disc and 
   the corona is non-thermal.
   
  {  In the LHS,  the radio flux is then
 positively correlated with the soft X-ray emission (3 -- 25 keV,
Corbel et al.  2000, 2003; Gallo, Fender, Pooley 2003). 
The X-ray emitting region and the jet seems to be intrinsically associated. 
This led to the now widely accepted idea that the corona (or hot thick disc) of the LHS 
constitutes the base of the jet (Fender et al. 1999; Merloni \& Fabian 2002; 
Markoff, Nowak \& Wilms 2005)}
 
\object{Cygnus~X-1} represents a prime target for the {\it INTEGRAL} mission (Winkler et
al. 2003) launched in 2002 October 17, whose instruments offer an unprecedented
spectral coverage at high-energy, ranging from 3 keV to several MeV.
\object{Cygnus~X-1} was extensively observed (1 Ms) during the calibration
phase of the mission (Bouchet et al. 2003, Pottschmidt et al. 2003,
Bazzano et al. 2003). At that time, the source presented all the
characteristics of the LHS. 
The source was later observed again during the galactic plane survey 
and open-time programme, and occasionally found 
in IMSs (Cadolle Bel et al., 2005, hereafter CB05). 
In this paper we focus on the results of the first 
observation of \object{Cygnus~X-1} in the open time programme. 
This  300 ks  observation was performed on 2003 June 7-11 (rev 79/80)
with a 5 $\times$ 5 dithering pattern\footnote{{\it INTEGRAL} observations 
are made of a succession of exposures of about 30 minute duration 
 with varied pointed directions to enable {\it SPI} image deconvolution. 
 Such a 30 minute pointing is called a science window}
 (the effective exposure time was 275 ks
for {\it JEM-X}2, 292 ks for {\it IBIS/ISGRI}, and 296 ks for {\it SPI}).
At this epoch, the {\it RXTE} All Sky Monitor count rate of \object{Cygnus~X-1} was
higher than in typical LHS by up to a factor of 4, and the
light curve showed strong X-ray activity characteristic of
state (or failed state) transitions 
(see Fig.~\ref{fig:asm}). We also combine the {\it INTEGRAL} data with
the results of coordinated radio observations (15 GHz) performed
with the Ryle telescope.   
In Sec.~\ref{sec:average}, we present a spectral analysis of the {\it JEM-X}, 
{\it IBIS/ISGRI} and {\it SPI} spectra averaged over the whole duration 
of the observation, in Sec.~\ref{sec:vari} we study 
the strong broad band variability of the source during the observation.

\begin{figure*}[!t]
\centering
 \resizebox{\textwidth}{!}{
\includegraphics[width=\linewidth]{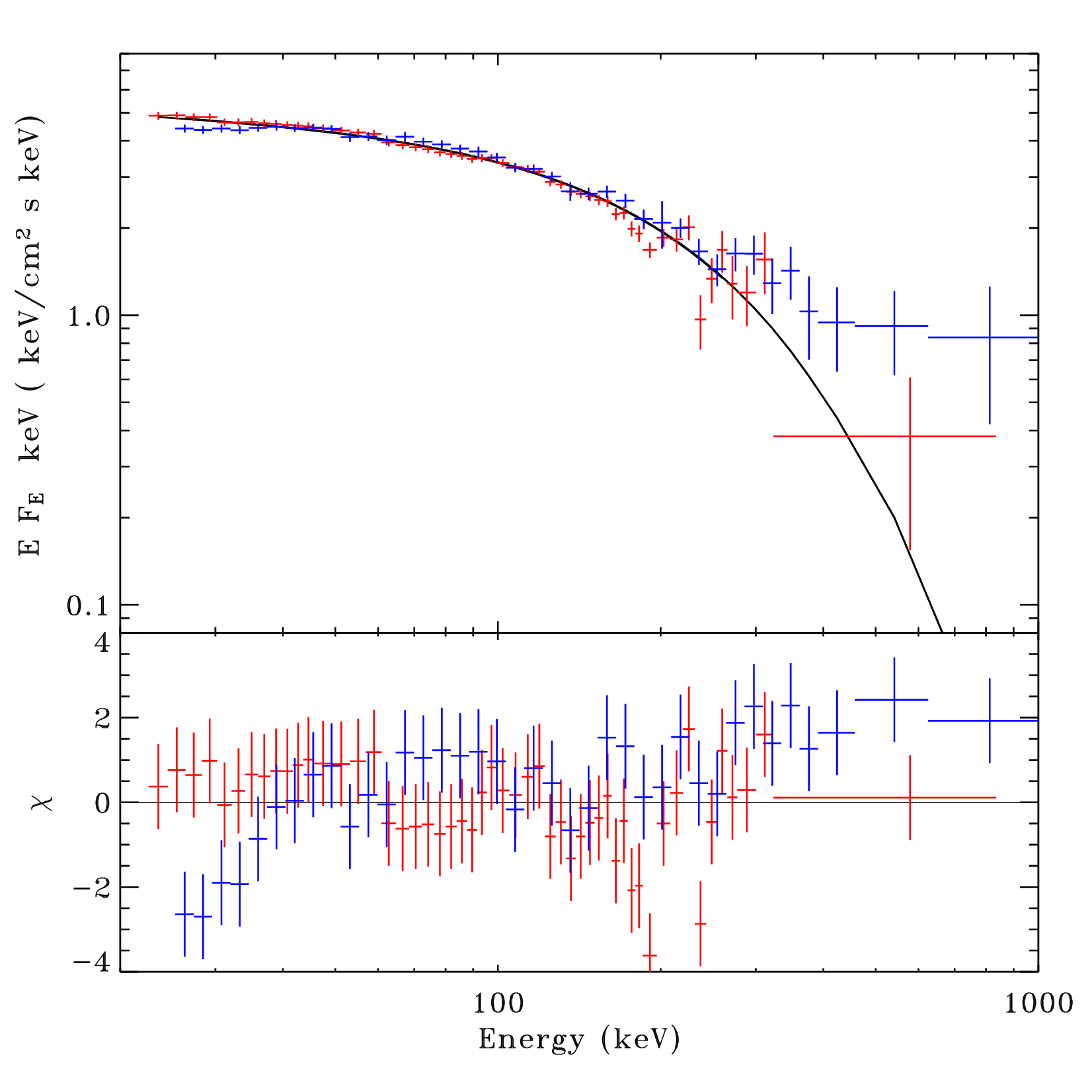}
\includegraphics[width=\linewidth]{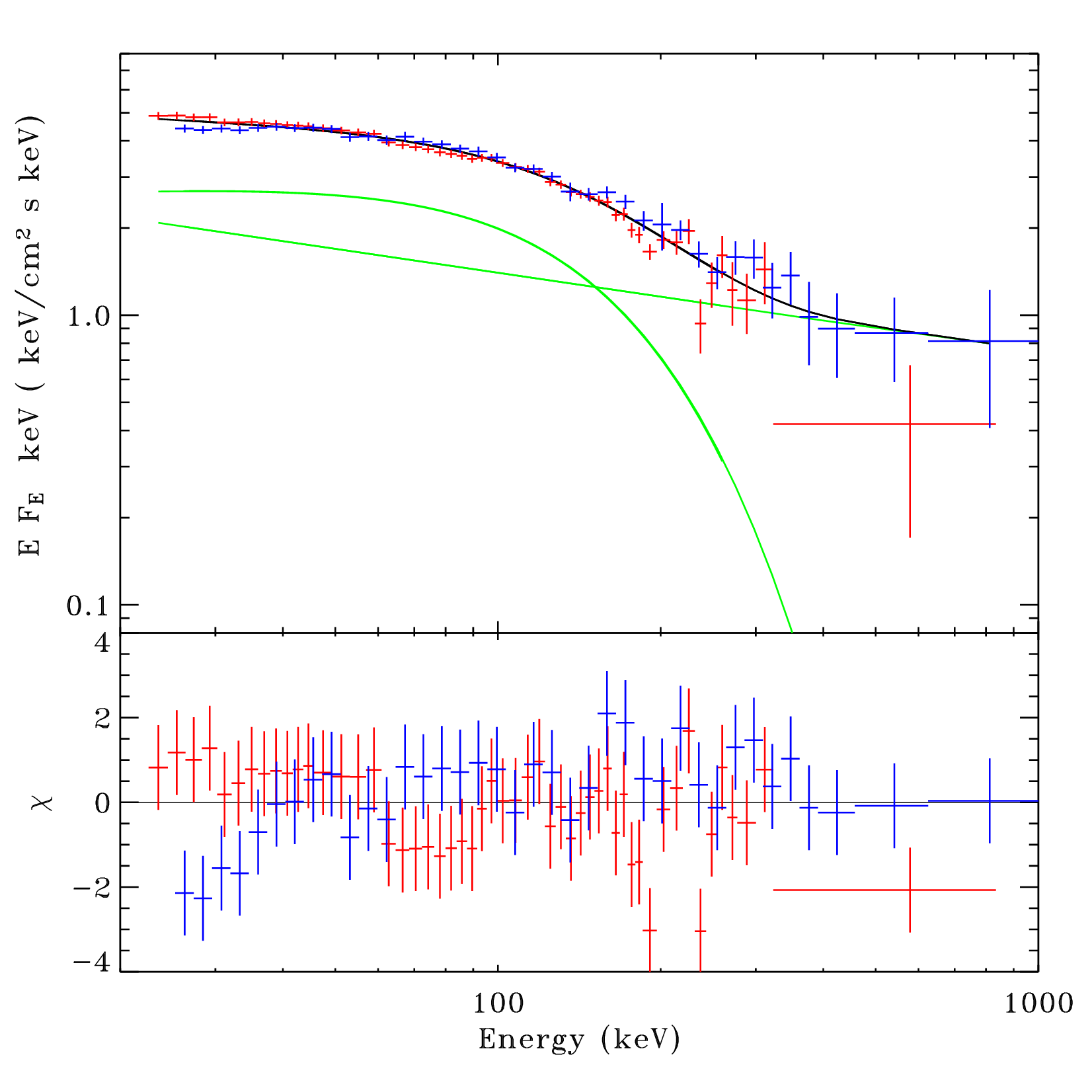}}
\caption{Best fits of the {\it IBIS/ISGRI} (red) and {\it SPI} (blue) data with a pure comptonization model ({\sc compps}, left) and
 Comptonisation plus power-law (right). \label{fig:tail}}
\end{figure*}

\begin{figure}[!t]
\centering
\includegraphics[width=\linewidth]{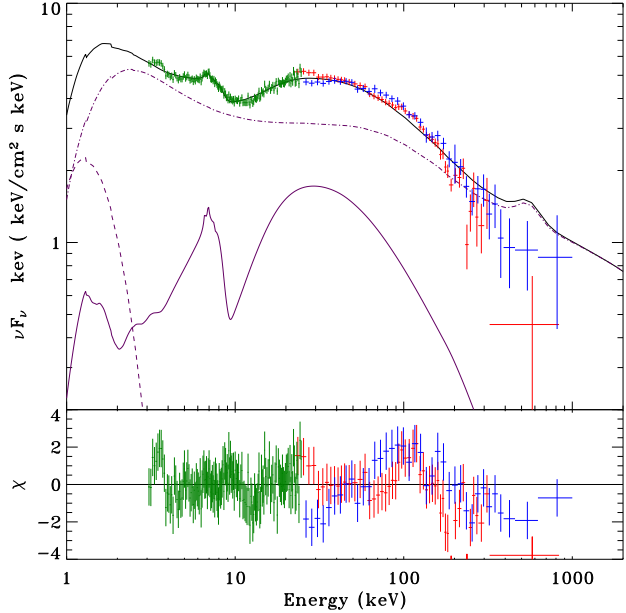}
\caption{Joint {\it JEM-X}/{\it SPI}/{\it ISGRI} spectrum of \object{Cygnus~X-1} 
averaged over revolutions 79 and 80. The data are fitted with the 
thermal/non-thermal hybrid Comptonisation model {\sc eqpair}
 with power-law injection of relativistic electrons (see text and table~\ref{tab:jointfit}). 
 The lighter curves show the reflection component (solid), 
 the disc thermal emission (dashed) and the Comptonised emission (dot-dash).
 The green, red and blue crosses show the {\it JEM-X},
  {IBIS/ISGRI} and {SPI} data respectively}
 \label{fig:spec7980} 
\end{figure}

\begin{figure}[!t]
\centering
\includegraphics[width=\linewidth]{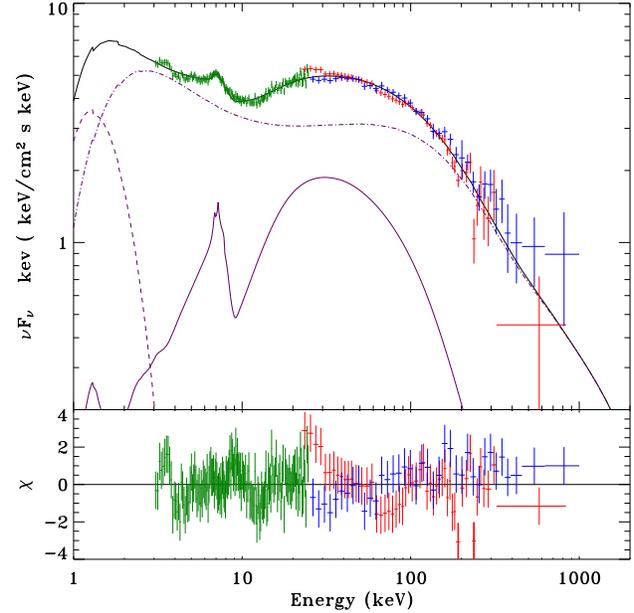}
\caption{Same as in Fig.~\ref{fig:spec7980} 
except that the fitting model is now {\sc eqpair} 
 with \emph{ mono-energetic} injection of relativistic electrons with Lorentz factor $\gamma_{\rm inj}=8.6$
 (see text and table~\ref{tab:jointfit}). 
\label{fig:spec7980mono} }
\end{figure}

\begin{figure}[!t]
\centering
\includegraphics[width=\linewidth]{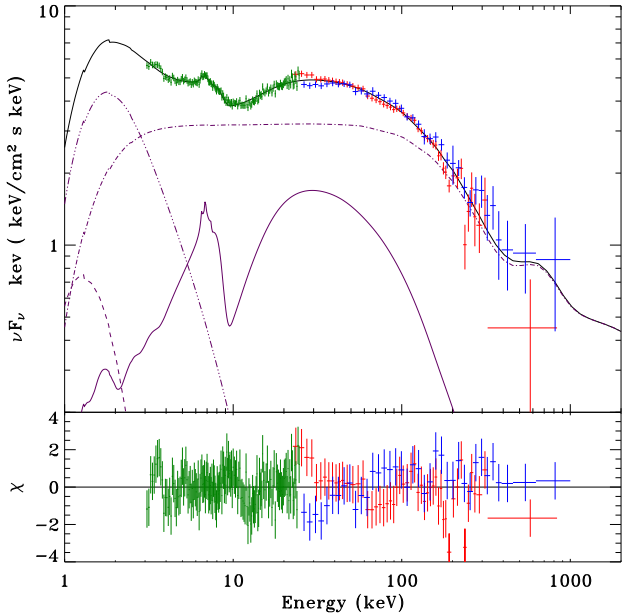}
\caption{Same as in Fig.~\ref{fig:spec7980}
except that the fitting model is now {\sc eqpair} with a power-law injection 
 relativistic electrons \emph{plus} an additional warm Comptonisation soft 
 component ({\sc comptt}, see text and table~\ref{tab:jointfit}). 
 The {\sc comptt} component is shown by the 3-dot-dashed curve}.
 \label{fig:spec7980comp} 
\end{figure}

\section{Average Spectrum}\label{sec:average}

\subsection{Data processing}

We reduced the {IBIS/ISGRI} and {\it JEM-X} data 
with the standard analysis procedure of the Off-Line Scientific Analysis {\sc OSA 4.2}. released by the ISDC, 
whose algorithms are described in Goldwurm et al. (2003) and Westergaard et al. (2003) for {\it IBIS} and 
{\it JEM-X} respectively. A basic selection was performed to exclude those pointings too close to radiation belt entry or exit or spoilt by too large noise.
 The {\sc SPI} data were preprocessed with {\sc OSA 4.2}  using the standard energy calibration gain coefficients per orbit and excluding bad quality pointings which have anomalous exposure and dead 
time values, or with a high final chi-squared during imaging. {\sc spiros 9.2} (Skinner \& Connell 2003)
 was used to extract the spectra of \object{Cygnus~X-1}, Cygnus X-3 and EXO 20390+375, 
with background model proportional to the saturating event count rates in the Ge detector.
Concerning the instrumental response, version 15 of the IRF (Image Response Files) and version 2 of the RMF (Redistribution Matrix Files) were used.
 
 We produced {\it JEM-X} (3-25 keV), {\it ISGRI} (20-800 keV) and {\it SPI} (25-1000 keV) 
 spectra averaged over revolution 79 and 80. Uncertainties of 3 \% were added in quadrature to
  all three spectra to account for systematic errors. The resulting spectra were fitted using {\sc xspec v11.3.1}

\subsection{{\it JEM-X} spectrum}

The best  power-law fit to the 3--20 keV {\it JEM-X} spectrum  has a spectral index
$\Gamma=2.16$ which is intermediate between LHS and HSS spectral indices.
The fit is however not statistically acceptable
($\chi^{2}/\nu=611/131$) with strong residuals indicating the
presence of strong reflection and broad iron line features and a soft component 
(see below in Sect.~\ref{sec:jointspec} and Fig.~\ref{fig:spec7980}).

\subsection{{\it SPI} and {\it IBIS/ISGRI} spectra}

The spectra show a highly significant evidence for a cut-off (or at least a break) 
around 100 keV (see Fig.~\ref{fig:tail}).
We first fit the {\it ISGRI} and {\it SPI} spectra independently with the Comptonisation 
model {\sc compps} (Poutanen \& Svensson 1996)
The {\sc compps} model provides a reasonable fit with reduced $\chi^{2}$ values 
0.82 and 1.12 for {\it ISGRI} and {\it SPI} respectively.
The best-fit parameters obtained for both instruments 
(shown in table~\ref{tab:fitsibispi})
 are compatible within the error bars. 
 
However the residuals show an excess emission with 
respect to the Comptonisation model above the thermal cut-off in the {\it SPI} spectrum,
which is not detected in the {\it ISGRI} spectrum. 
This prompted us to reprocess the {\it SPI} data 
 using different background models based on 
 an empty field observation to model the uniformity map of the detection plan.
 We also tested using the instrument team
 processing software instead of OSA. This did not affect the presence of a high 
 energy excess in the {\it SPI} spectrum.
 We concluded that the non detection of this excess by {\it ISGRI} 
 was attributable to its poorer sensitivity above 200 keV. This is also the reason why 
 the {\it ISGRI} fit appears statistically better. 
 Within the errors bars the highest energy IBIS and SPI data points 
 are not too far and still compatible.

  To obtain better constraints on the parameters we then fit simultaneously 
  the {\it SPI} and {\it ISGRI} data.
  The results are shown in Fig.~\ref{fig:tail}.
   In all fits we allow for a free normalisation constant of the model for each instrument 
   to correct for inter-calibration errors.
  The resulting difference in normalisation between the instruments
   never exceeds 20 \%.
  The simple {\sc compps} model leads to best-fit 
  parameters that are intermediate between those obtained
  for the individual {\it ISGRI} and {\it SPI} spectra, but the statistical quality of the fit is
   not as good  ($\chi^{2}/\nu=1.4$). { As a further test for the presence of a high-energy excess,
 we refit both spectra with {\sc compps} adding  a power law component.}
  When we allowed for this additional component the reduced $\chi^{2}$ 
  decreased to 1.05. This improvement is highly significant, a F-test 
  ($\Delta\chi^2=36$ for 2 additional parameters) 
  shows that the probability that this improvement
   occurred by chance is  5 $\times 10^{-7}$.
  We conclude that either the high-energy cut-off is not due to thermal
   Comptonisation (e.g. associated to a cut-off in a non-thermal lepton distribution)
  or there is an additional non-thermal component at high-energy.
  We stress that the presence of the high energy excess is not in conflict 
  with the {\sc ISGRI} data although this instrument was not able to detect it.

\begin{table}
\caption{Best-fit parameters  of the {\it SPI} and {\it IBIS} spectra
 fitted with thermal-Comptonisation and thermal-Comptonisation plus power-law models.
For each instrument the first line gives the results of the {\sc compps} model alone
 (Comptonisation tamperature $kT$ and Thomson depth $\tau$)
while the second line
gives  the results of the fit with {\sc compps} plus a power-law (photon index $\Gamma$).
In the {\sc compps} model the black body temperature 
of the soft seed photons was fixed at 0.1 keV in all fits.}       
\label{tab:fitsibispi}      
\centering          
\begin{tabular}{c c c c c c c}     
\hline\hline      
 data     & $kT$  (keV)                     & $\tau$                             &  $\Gamma$                &  $\chi^2/\nu$\\ 
\hline                    
 {\it ISGRI}       & $79^{+7.5}_{-7.3}$ & $1.06^{+0.17}_{-0.14}$   &                                              & $46/56$\\
 {\it ISGRI}       & $51^{+12}_{-12}$   &  $2.01^{+0.59}_{-0.81}$   &        $2.39^{+0.14}_{-1.57}$            &  43/54\\
  {\it SPI}         & $90^{+18}_{-8}$    & $0.98^{+0.15}_{-0.24}$   &                                                &  50/44 \\
  {\it SPI}         &  $45^{+15}_{-12}$  & $2.38^{+1.1}_{-0.49}$    &      $2.06^{+0.24}_{-0.70}$         & 23/42 \\
  {\it SPI}+{\it ISGRI} & $84^{+6.3}_{-6.8}$ & $1.00^{+0.14}_{-0.10}$ &                               & 142/102\\  
  {\it SPI}+{\it ISGRI} & $42^{+15}_{-4}$      & $2.64^{+0.50}_{-0.33}$  &    $2.27^{+0.66}_{-0.77}$  & 106/100\\
\hline                  
\end{tabular}
\end{table}

\begin{table}
\caption{Best-fit parameters of the joint {\it JEM-X}, {\it SPI} and {\it IBIS/ISGRI}
 spectra with hybrid thermal/non-thermal Comptonisation models ({\sc eqpair}, 
 see Sect.~\ref{sec:jointspec}) .
 The  temperature of the inner  disk ({\sc diskpn}) was fixed to $kT_{\rm max}$=0.3 keV
  in all fits. The soft photon compactness is fixed at $l_{\rm s}=10$. 
 The absorbing column density is $N_{\rm h}=5 \times 10^{21}$ and the inclination 
 angle 45 degrees. The fit parameter $\tau_{\rm p}$
  refers to the Thomson optical depth of the electrons associated with ions. 
 The table also  gives the resulting 
 total optical depth   $\tau_{\rm T}$, including electron-positron
   pairs whose production is calculated self-consistently, and the temperature $kT_{\rm e}$
    of the thermalised particles, computed according to energy balance.
    The extrapolated  0.1-1000 keV  model flux,  $F_{\rm bol}$, and thermal disc component flux,  $F_{\rm disc}$, are given 
  in units of $10^{-8}$~ergs~s$^{-1}$~cm$^{-2}$.
     The first column shows the results of the fit with a power law injection of non-thermal 
  electrons with Lorentz factors ranging from $\gamma_{\rm min}$=1.3 to
   $\gamma_{\rm max}$=1000. The data are then fitted for the index of the distribution of
   injected electrons $\Gamma_{p}$.
  The second column shows the best fit parameters for a mono-energetic injection 
  of electrons, the fit parameter $\gamma_{\rm inj}$ represents the Lorentz factor of the injected particles.
  Finally, the third column gives the results for a power-law injection plus an additional 
  warm Comptonisation component ({\sc comptt}).
  Its   best fit temperature $kT_{\rm comptt}$, Thomson depth $\tau_{\rm comptt}$ and 
  flux $F_{\rm comptt}$ are also shown. In this fit $\Gamma_{p}$ pegged to its minimum allowed boundary (2)}             
\label{tab:jointfit}      
\centering          
\begin{tabular}{c c c c}     
\hline\hline       
      Model                   &   {\sc eqpair}  pow.                      & {\sc eqpair} mono.                     &     {\sc eqpair}+{\sc comptt}     \\    
      \hline
 $l_{\rm h}/l_{\rm s}$ &   $1.19^{+0.07}_{-0.06}$    & $0.85^{+0.02}_{-0.03}$     &  $3.06^{+0.41}_{-0.16}$                     \\ 
 $l_{\rm nth}/l_{h}$    &       $1^{+0}_{-0.04}$         &   $0.51^{+0.04}_{-0.04}$   &        $0.23^{+0.37}_{-0.02}$                \\    
 $\tau_{\rm p}$           &     $1.31^{+0.02}_{-0.04}$  &    $0.55^{+0.01}_{-0.06}$  &  $0.79^{+0.13}_{-0.07}$                       \\  
 $\Gamma_{\rm p}$ or $\gamma_{\rm inj}$ & $2.68^{+0.02}_{-0.01}$  &  $8.41^{+0.62}_{-0.92}$ &  $2^{+0.21}_{-0}$\\                                            
  $\Omega/2\pi$              &       $0.67^{+0.09}_{-0.02}$ &   $0.71^{+0.09}_{-0.03}$   &    $0.62^{+0.09}_{-0.06}$            \\   
  $\xi$ (erg cm s$^{-1}$) &        $2063^{+463}_{-367}$     &    $525^{+143}_{-84}$   &     $991^{+297}_{-176}$                          \\
  $E_{\rm line}$ (keV)    &   $6.76^{+0.36}_{-0.33}$        &     $7.02^{+0.32}_{-0.23}$  & $6.67^{+0.25}_{-0.25}$                         \\    
      EW (eV)                     &    $55^{+33}_{-34}$            &           $90^{+38}_{-24}$                 &    $76.3^{+34}_{-29}$  \\ 
  $kT_{\rm {\sc comptt}} $ (keV) &                                   &                            &         $2.53^{+0.20}_{-0.99}$       \\ 
  $\tau_{\rm {\sc comptt}}$        &                                        &                           &     $2.73^{+0.34}_{-0.27}$   \\
   $F_{\rm comptt}$                                                    &                     &                               &      1.52         \\
   $F_{\rm disc}$                                                        &   1.9             &         3.0                    &     0.63         \\
   $F_{\rm bol}$   &   7.5            &         7.9                       &    5.9       \\
   $kT_{\rm e}$ (keV)      &        20.3                                &   50                               &            65      \\ 
  $\tau_{\rm T}$            &  1.36                                      &              0.55                 &     1.01      \\

  $\chi^2$/d.o.f.             &       316/245                            &        244/245                  &  216/242   \\  
\hline                  
\end{tabular}
\end{table}

\subsection{Joint spectrum}\label{sec:jointspec} 

We now consider the joint {\it JEM-X}/{\it ISGRI}/{\it SPI} spectrum. The simple models
 considered in the previous section 
do not provide an acceptable description of the broad band spectrum: 
strong reflection features and a soft-excess are required by the {\it JEM-X} data.
CB05 present spectral fits of the same data with various thermal and non-thermal Comptonisation models. 
In these models
the soft excess in the JEM-X spectrum is accounted for by thermal emission of the accretion disc.
 However, although, these fits are statistically acceptable, they require high temperatures of 
 the accretion disc (1.21$\pm$0.29 keV).
    As mentioned  in CB05 such a high temperature of the accretion
    disc is not realistic. Since the distance and scale of the accretion disc (i.e. mass of the
     black hole) are quite well known for this source, the intrinsic luminosity 
     of the accretion disc is constrained: if the accretion disc in \object{Cygnus~X-1} had
    an inner temperature of 1 keV, it 
    would produce a flux at least one order of magnitude larger than what is observed.

  In order to get a more physically motivated representation of the data,
 we investigate in more details thermal/non-thermal hybrid Comptonisation models and
 attempt to fit the data using different variants of the {\sc eqpair} model (Coppi 1999; G99; 
 Frontera et al. 2001 (hereafter F01); Zdziarski et al. 2002, 2004) 
 where the accretion disc temperature is arbitrarily fixed to a reasonable value.
 { This model was shown to be
   successful in accounting
   for the high energy spectra of \object{Cygnus~X-1} and other black 
   holes candidates in different spectral states and 
  over a very broad energy band ranging from soft X-rays to gamma-rays
  ( see e.g. McConnell et al. 2000, 2002). 
  The wide use of this model in the literature will allow us to compare our observation 
  with previously published results.}

\subsubsection{The hybrid thermal/non-thermal Comptonisation model ({\sc eqpair})}\label{sec:eqpair}

A detailed description of the {\sc eqpair} model can be found in G99.
Its main ingredient is a spherical hot plasma cloud with
continuous acceleration of electrons intended to model the emission of the hot disc/coronna.
The high-energy electrons lose energy because of Compton,
Coulomb, and bremsstrahlung processes and thus establish
a steady-state distribution. 
 At high energies, the distribution
is non-thermal (power-law like), but at low energies a
Maxwellian distribution is established.
 The temperature of
the Maxwellian population, $kT_{\rm e}$ is determined by balance between
Compton gains and losses, Coulomb heating by high energy
electrons, bremsstrahlung losses, and direct
heating (e.g., Coulomb heating by energetic ions). The total
number of electrons (not including $e^{+}-e^{-}$ pairs, the
production of which is also taken into account) is determined
by the corresponding Thomson optical depth $\tau_{\rm p}$ which is a free parameter. 
The cloud is illuminated by soft thermal 
photons emitted by an accretion disk. These photons
serve as seed for Compton scattering by both thermal and
nonthermal electrons.
The system is characterised by the power $L_{i}$ supplied to its
different components. We express each of them dimensionlessly
as a compactness, $l_{i}= L_{i} \sigma_{T}/ (\mathcal{R}m_{e}c^3)$.  
$\mathcal{R}$ is the characteristic dimension of the plasma,  $\sigma_{T}$ 
is the Thomson cross-section. $l_{s}$, $l_{th}$, $l_{nth}$, and $l_{h}$= $l_{th}$+$l_{nth}$
 correspond to the power in soft disk photons entering the plasma,
  thermal electron heating, electron acceleration and the total power supplied to the plasma.
We follow G99 and fixed  $l_{s}=10$.

The disc spectrum incident on the plasma is modelled as
coming from a pseudo-Newtonian accretion disk extending
from $R_{out}=10^{3}R_{g}$ down to the minimum stable orbit, $R_{in}=6R_{g}$.
Its spectral shape is then characterised by the
maximum colour temperature of the disk, $kT_{max}$. 
Previous observations of \object{Cygnus~X-1} with X-rays telescopes indicate temperatures 
ranging from 0.1 keV in the LHS to up to 0.6 keV in the HSS (G99;F01). 
For such temperatures the peak of the thermal disk emission is below 
the energy range covered by {\it INTEGRAL}, and the disk temperature  cannot 
be constrained by our observations,
 we therefore fixed $kT_{max}=0.3$ keV.
The covering factor of the corona is unity.

 The spectrum from both reflection and the Fe $K_{\alpha}$ is
calculated, taking into account relativistic smearing with 
the emissivity dependence $\propto R^{-2}$. 
The reflecting material is allowed to be
ionised, with the degree of ionisation characterised by the
ionisation parameter $\xi$. 
We follow G99 and fixed the column density of absorbing material along the line of sight
 to $N_{H}=5 \times 10^{21}$ cm$^{-2}$ and the inclination angle of the system at  45 degrees.

\subsubsection{Fit results}

We first performed a fit similar to those of G99 and F01 with the non-thermal electrons 
injected with a power-law distribution of Lorentz factors ranging from $\gamma_{\rm min}=1.3$ to  
$\gamma_{\rm max}=1000$. The upper and lower limits $\gamma_{\rm min}$ and $\gamma_{\rm max}$ 
where kept fixed while fitting for the power-law index $\Gamma_{\rm p}$. 
 This results in an acceptable fit 
with a reduced $\chi^2$ of 1.29. The unfolded broadband spectrum and residuals are
 shown in Fig.~\ref{fig:spec7980}. The  best-fit parameters 
are presented in Table~\ref{tab:jointfit}. 
$l_{\rm h}/l_{\rm s}$ is about unity i.e. intermediate between 
what is generally found in the LHS (4--10, see Ibragimov et al. 2005) and the HSS
 ($\lesssim 0.4$), the heating of the plasma is dominated 
 by the non-thermal acceleration ($l_{\rm nth}/l_{\rm h}=1$).
 Overall the parameters are similar to those obtained 
 by G99 for the IMS of 1996 May 23.
 However an inspection of the residuals (Fig.~\ref{fig:spec7980}) 
 shows that the high energy part of the {\it ISGRI}/{\it SPI} spectrum 
  is not well represented by this model. 
The model does not provide a good description of the shape 
 of the high-energy cut-off and overestimates the measured flux above 200 keV.
{ As an alternative we attempted to fit the spectrum assuming that the electrons are injected 
 at a single Lorentz factor $\gamma_{\rm inj}$ instead of a power-law distribution.}
   Such a mono-energetic injection is not 
 expected in the case of shock acceleration but could be achieved 
  in reconnection events that are expected to power the corona.
  The resulting fit is displayed on Fig.~\ref{fig:spec7980mono}
   and the unabsorbed best fit model spectrum is shown on Fig.~\ref{fig:interp}.
   The best-fit parameters  are shown on Table~\ref{tab:jointfit}. Note that the fraction of non-thermal power 
  is now only $l_{\rm nth}/l_{\rm h}=0.51$. This model gives a good description
   of the {\it INTEGRAL} spectrum. 
   The best-fit value  for the maximum Lorentz factor of the electrons
   $\gamma_{\rm inj}=8.6$ implies that the once scattered soft photons 
    reach a maximum energy of about $4\gamma_{\rm inj}^{2}kT_{\rm max}\sim$ 90 keV. In other words, in this model 
    the high-energy cut-off observed in the spectrum
     is non-thermal and corresponds to the upper end of the non-thermal electron distribution.
 
  So far we have shown that the soft excess in the JEM-X data does not require 
  an unphysical high accretion disc temperature and can be accounted for by an hybrid 
  particle distribution.  In this context, our data favour mono-energetic injection over
   power-law acceleration.  However, this is not the only possible interpretation.
 Indeed, such a soft excess is not uncommon
     in IMSs and the softest LHSs.
     It is often interpreted as a hot spot on the accretion disc or alternatively 
     as a component due  Comptonisation in the warm upper layers of the disc.
      It can be accounted for by adding 
     a second Comptonisation component with a temperature of a few keV
     (Di Salvo et al 2001; F01; Zicky, Done \& Smith 2001). 
    { Alternatively, Markoff, Nowak \& Wilms (2005), suggest
     that this soft component originates from optically thin synchrotron emission in the jet.}
      
     To check the effect of such an additional soft component on the derived parameters of the 
     {\sc eqpair} model,  we fit the spectrum with a model consisting 
  in the power-law injection {\sc eqpair} model, including the same 0.3 keV 
  thermal component,  relativistically smeared iron line and ionised reflection component 
  as in our previous models (see Sect.~\ref{sec:eqpair})
  plus an additional Comptonisation component ({\sc comptt} model in XSPEC) 
  in a warm medium. 
     
  This led to an excellent fit of the {\it INTEGRAL} data,  with $\chi^2/\nu=0.89$, that is shown in 
  Fig.~\ref{fig:spec7980comp} and Table~\ref{tab:jointfit}. 
  The additional Comptonisation component has a best-fit temperature of
   2.5 keV and a Thomson optical depth of 2.7.
   The fit is much better than what obtained 
   with the powerlaw injection {\sc eqpair} model. 
  The probability that the improvement 
  obtained when the soft {\sc comptt} component is added to the powerlaw injection
   {\sc eqpair } model  occured by chance is 7$\times$ $10^{-20}$
    according to F-test ($\Delta \chi^2 =100$ for 3 additional parameters).
    { This fit is statistically comparable to that obtained with mono-energetic injection.
    It demonstrates that mono-energetic injection is far from inevitable.

    We also note an interesting difference between 
    power-law and mono-energetic injection models: the presence of a broad 
    $e^{+}-e^{-}$ pair annihilation line around 511 keV that is clearly visible in Fig.~\ref{fig:spec7980}
      and ~\ref{fig:spec7980comp} (power-law injection) but absent 
      from Fig.~\ref{fig:spec7980mono} (mono-energetic injection). 
      In the power-law model some electrons are injected at very high Lorentz factors 
      ($\lesssim 1000$) these electrons efficiently up-scatter soft-photons at energies 
      above the pair production threshold, leading to a pair cascade and subsequent annihilation, 
      whereas in the mono-energetic case the pair content of the plasma is negligible 
      due to the much lower energy of the electrons  ($\leqslant$ 8.41). 
      Unfortunately the poor statistics of the data does not allow us to discriminate 
      between these models on the basis of the annihilation line.} 
         
   In the {\sc eqpair+comptt} spectral model the fraction of non-thermal power 
   ($l_{\rm nth}/l_{\rm h}$) is reduced 
   to about 20 \% (to be compared to 100 \% and 50 \% obtained respectively
   for the simple power-law  and mono-energetic injection models).
   This shows that the measured non-thermal fraction is sensitive to the assumptions
   of the spectral model. 
    In the latter fit the hot plasma is essentially thermal, in agreement with  the conclusions of
    CB05. Indeed, the equilibrium temperature and optical depth we obtain are similar 
   to what obtained by these authors fitting the same data with 
   a simple thermal Comptonisation model plus a multicolour disc spectrum. 
   We also note  that in the {\sc eqpair+comptt} fit, the hard to soft compactness ratio is larger ($l_{\rm h}/l_{\rm s}\simeq 3$ instead of $\simeq 1$ in the previous fits). 
    Overall the parameters are closer to what is usually obtained in the LHS.
   
    In short, our spectral analysis does not favour the simple power-law injection {\sc eqpair}
     model. Rather, models  including an additional soft component (and a weak non-thermal
      fraction),
     or alternatively, models with an important thermal fraction but mono-energetic injection
       are preferred. 
     There are however some caveats in our spectral analysis that might affect 
     any conclusion drawn from our fits:
     First, there are still open calibration issues, in particular
      regarding the X-ray monitor {\it JEM-X} specially below 5 keV. 
       It is in this energy range that the soft component shows up 
     and it is very important to constrain 
     the parameters of the hybrid model.  
     Second, our model assumes a unique emitting zone although 
     it is likely that in an IMS the situation is more complex.
    Indeed, the thermal hot flow of the LHS probably coexists
      with the non-thermal corona of the HSS in distinct regions of the accretion flow.
      If various emitting regions, with very different
       spectra and physical parameters, contribute to the observed spectrum, then the derived
       best-fit parameters have a poor physical significance.
      Similarly, and perhaps more importantly, the parameters determining the shape 
      of the spectrum vary not only within the accretion flow but also with time.
       As we will show below, the source exhibited a strong spectral variability during 
      the observation  with the appearance of spectra ranging from quasi-LHS  
      to quasi-HSS. Under such circumstances,  
      the precise value of the best-fit parameters of the average spectrum may be physically 
      irrelevant.
      
\section{Spectral Variability}\label{sec:vari}

\subsection{Set up}

In order to study the spectral variability of the source during the
observation, we produced light curves in 16 energy 
bands ranging from 3 to 200 keV\footnote{namely: 3--4, 4--5, 5--6, 6--7, 7--9, 9--11, 11--13, 13--15,
15--20, 20--30, 30--40, 40--50, 50--80, 80--100, 100--140 and 140--200 keV}
 with a time resolution of the duration of a science
window (i.e. $\sim$ 30 min). 
Above 200 keV, the variability is dominated by statistical noise.
The count rate in each band was then
renormalised so that its time average matches the energy flux
calculated from the best-fit model of the
joint average {\it JEM-X}/{\it ISGRI}/{\it SPI} spectrum 
{ shown in Fig.~\ref{fig:spec7980comp}}. 
{ Namely, for each energy band we compute the quantity
\begin{equation}
F(t)=\frac{\mathcal{C}(t)}{\bar{\mathcal{C}}} \bar{F},
\end{equation}
where $\mathcal{C}(t)$ is the mean count rate during pointing $t$, 
$\bar{\mathcal{C}}$ is the count rate averaged over the whole observation, 
 $\bar{F}$ is the observation average energy flux in this band given by the best fit model.
 We use $F(t)$ as a proxy for the instantaneous energy flux, therefore neglecting the effects
  of the spectral variations on the instrumental response.}
  This simple deconvolution method provides us with a convenient approximate of the
broadband energy spectrum for each pointing that will be useful for a
physical interpretation of the variability.
This method is much more convenient than fitting the spectra for each science window, 
in particular if one considers the large number of parameters required to fit the data and 
the poor statistic in the short exposure spectra.
On the other hand our method for estimating the spectra from the light curves
 also enables us to improve the photon statistics 
by  combining the {\it IBIS/ISGRI} and {\it SPI} instruments in the
energy range where they overlap. 
The time dependent flux $F(t)$ in an overlapping band is estimated as follows:
\begin{equation}
F(t)= \frac{\sigma_{\rm S}^2 F_{\rm I}(t) + \sigma_{\rm I}^2 F_{\rm S}(t)}{\sigma_{\rm S}^{2}+\sigma_{\rm I}^{2}}
\end{equation} 
Where $F_{\rm I}$ and $F_{\rm S}$ are the measured {\it ISGRI} and {\it SPI} fluxes in that
band; $\sigma_{\rm I}$ and  $\sigma_{\rm S}$ are their time averaged
uncertainties. This combination minimises the average uncertainty on
$F$.

\begin{figure}
   \centering
   \includegraphics[width=\linewidth]{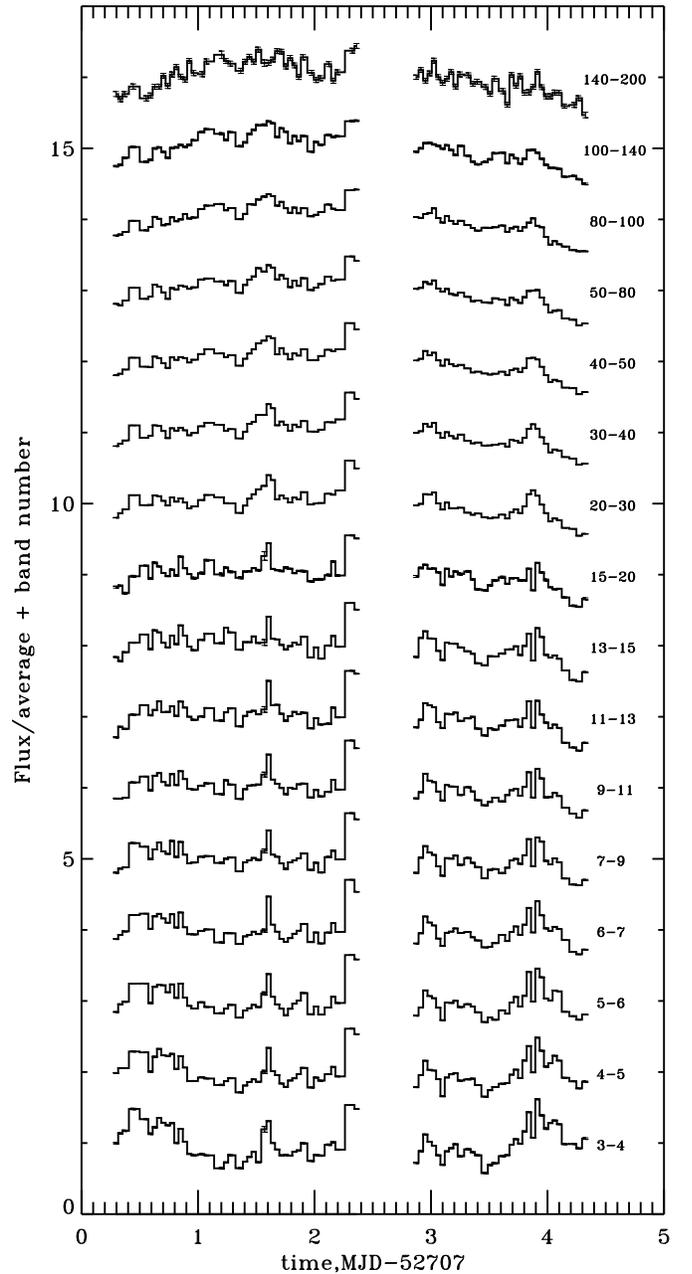}
      \caption{Light curves in different {\it JEM-X} an {\it ISGRI}/{\it SPI} bands
       with increasing photon energy from bottom to top, as labelled in keV. 
Each light curve was re-scaled to the average flux and then incremented
according to its energy range for clarity.}
         \label{fig:lcpca}
   \end{figure}

The resulting light curves are shown in Fig.~\ref{fig:lcpca}.
The time averaged 3--200 keV model flux is 
$\bar{F}_{3-200}=2.87 \times 10^{-8}$ ergs cm$^{-2}$ s$^{-1}$.
The energy flux has a rms amplitude  of 16 \%, and the ratio of the maximum
 to the minimum luminosity is 2.6.

   \begin{figure}
   \centering
   \includegraphics[width=\linewidth]{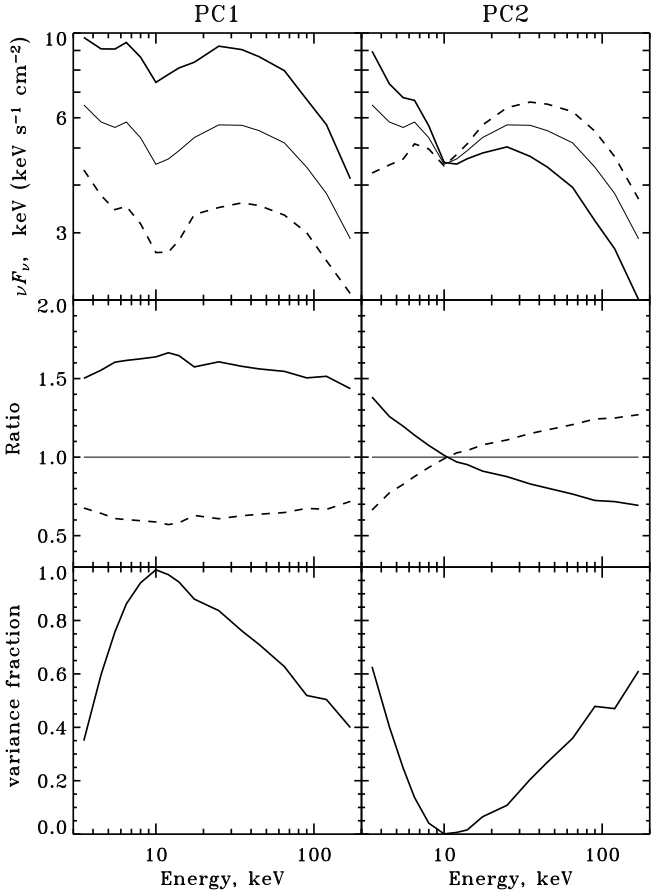}
      \caption{The 2 first principal components of variability.  
      The upper panels illustrate the effects of the each component on the shape 
      and normalisation 
      of the spectrum: time average spectrum (light line) 
      and spectra obtained for the maximum (thick solid line) and minimum (dashed line) 
      observed values of the normalisation parameter.
      The middle panels show the ratio of spectra obtained of the maximum and 
      minimum spectra to the average one.
      The bottom panels show the contribution of each component to the total
       variance as a function of energy.
              }
         \label{fig:pcatot}
   \end{figure}

    \begin{figure}
   \centering
 \includegraphics[width=\linewidth]{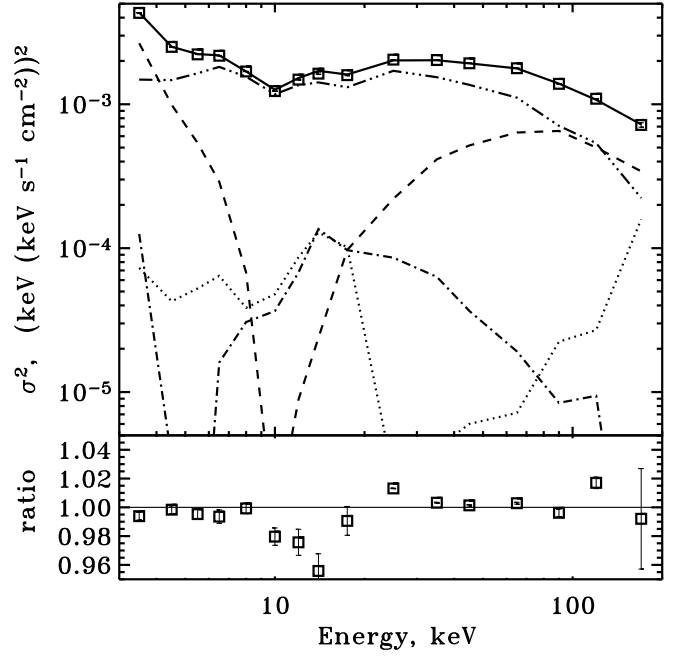}
      \caption{Observed variance spectrum (squares).
       The curves show the contribution of PC 1 (3 dots-dash),
       PC 2 (dashes), PC 3 (dot-dash),
       estimated statistical noise (dots) and their sum (solid). 
      The bottom panel shows ratio of the contribution 
      of PC~1+PC~2+PC~3+noise to the total observed variance 
      (values exceeding unity might be due to overestimated statistical noise)}
         \label{fig:pcarms}
   \end{figure}
   
       \begin{figure}
   \centering
 \includegraphics[width=\linewidth]{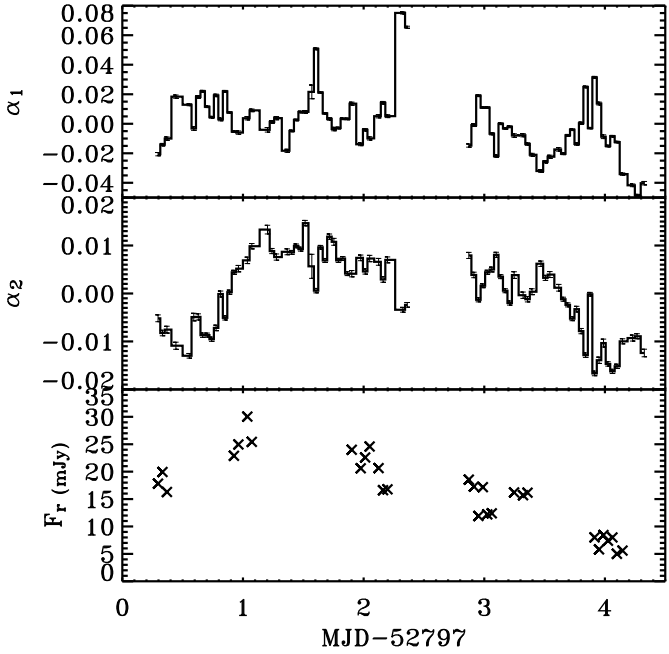}
      \caption{ Evolution of the parameters associated to PC 1 (top),
PC 2 (middle) and radio light curve (bottom) during the observation}
         \label{fig:pcavar}
   \end{figure}
\begin{figure*}
   \centering
 \resizebox{\textwidth}{!}{\includegraphics{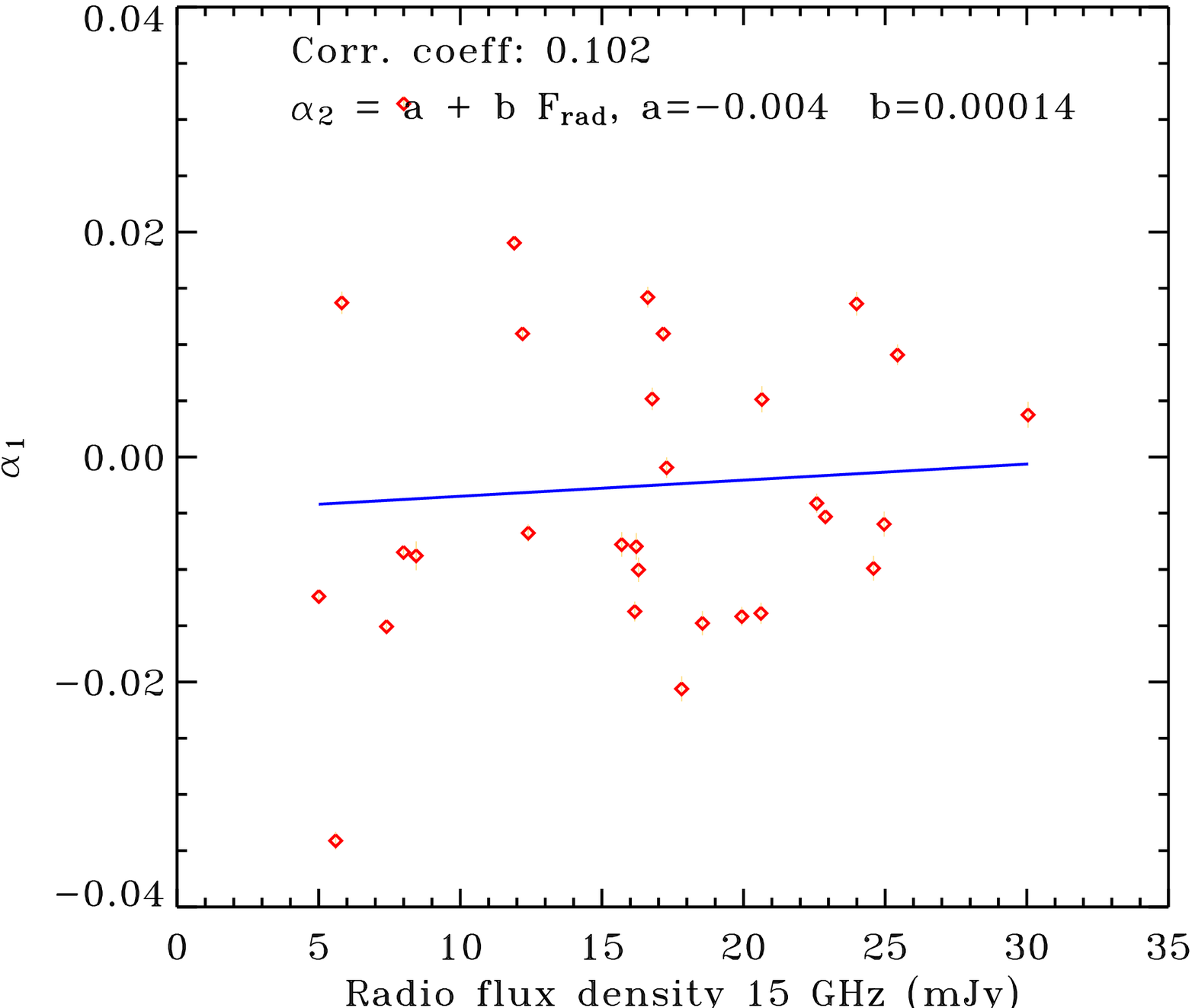}\includegraphics{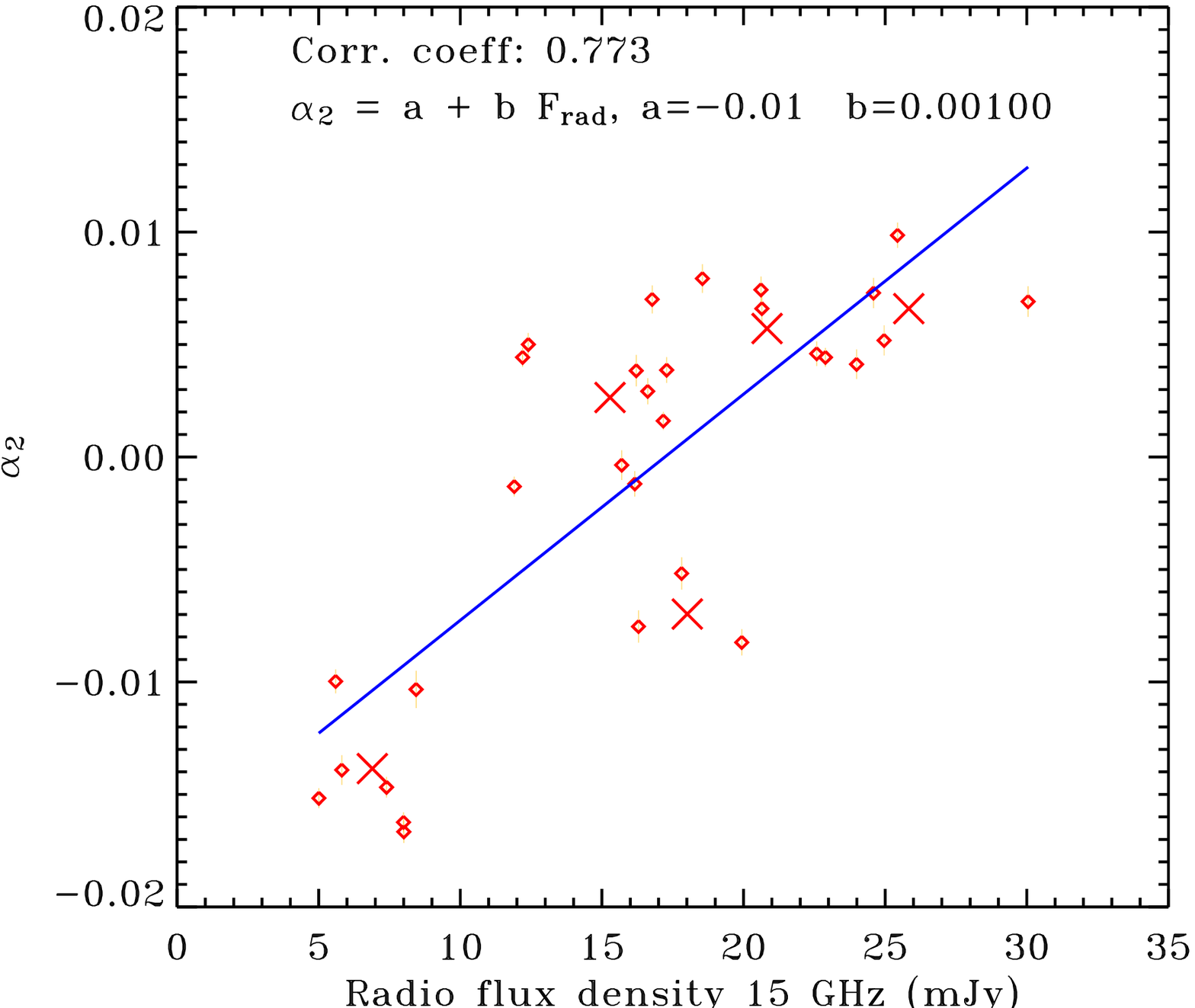}}
      \caption{PCA parameters $\alpha_{1}$ (left panel) and $\alpha_{2}$ (right panel) 
      as a function of the radio flux (diamonds). In both panels, the best linear fits are shown by the solid lines. 
The crosses indicate the time average over each of
the five periods of nearly continuous radio coverage 
 (see Fig ~\ref{fig:pcavar}).  While there is no convincing correlation between the
  radio flux and $\alpha_{1}$, the radio flux is correlated 
  to $\alpha_{2}$ at highly significant level.}  

         \label{fig:radal}
   \end{figure*}
   
\subsection{Principal Component Analysis (PCA)}

The light curves shown in Fig.~\ref{fig:lcpca} exhibit a complex 
and strong broad band variability of the spectra as well as the overall flux.
We use a principal component analysis (PCA) to seek for variability
patterns in our sample. PCA is a powerful tool for multivariate data analysis 
used for a broad range of applications in natural as well as social sciences 
(see e.g. Kendall, 1980).  It has also been used in astronomy.
For instance, Francis \& Wills (1999) provide a brief introduction to PCA as applied to quasar spectra.
 Previous application of PCA to spectral variability include 
Mittaz, Penston \& Snijders (1990) who discussed the UV 
variability of NGC 4151 and Vaughan \& Fabian (2004)
 for the X-ray variability of MCG6-30-15.
 
The main use of PCA is to reduce the dimensionality 
of a data set while retaining as much information as possible. 
It transforms a number of (possibly) correlated variables into a (smaller) 
number of uncorrelated variables called principal components.
 These principal components may define patterns
or correlations in the data set that can often be interpreted more easily
 than the original (large) data sets.

In our specific application we have $p=79$ spectra measured at times $t_{1}$, $t_{2}$, ..,,$t_{p}$ 
and bined into $n=16$ bins corresponding to energies $E_{1}$, $E_{2}$,...,$E_{n}$. 
{  Each spectrum can be thought of as a point in a $n$-dimensional space, so that
 the coordinate of the point $j$ along the  $E_{i}$ axis is given by the corresponding 
 energy flux $F(t_{j},E_{i})$.
 The shape and extension of the cluster of points formed by the whole set of spectra 
 characterises the source variability.
 The idea behind PCA is to determine a new coordinate system
  in which the description of this cluster will be simpler.

For this purpose, the data can be viewed as the  $p \times n$ 
matrix which coefficients are given by the energy fluxes $F(t_{j},E_{i})$. 
Then one can compute the $n \times n$ covariance 
matrix of the data.
In practice, PCA consists in diagonalizing this covariance matrix to obtain its
 eigenvalues and eigenvectors.
This procedure gives the coordinates of each eigen vector i.e. the  
$C_{k}(E_{i})$ coefficients denoting 
the coordinate of  the $k$-th eigenvector along  $E_{i}$ axis.

These eigenvectors define a new coordinate
system, in this $n$-dimensional parameter space, which best
describes the variance in the data. The first principal component, or
PC 1 (the eigenvector with the highest eigenvalue), marks the direction
through the parameter space with the largest variance . The next
Principal Component (PC 2) marks the direction with the second
largest amount of variance, etc...

Let $\alpha_{k}(t_{j})$ be the new coordinate of the spectrum $j$ along the $k$-th eigen 
vector, then the relation between the two coordinate systems can be written as follows:  
\begin{equation}
F(t_{j},E_{i})=\bar{F}(E_{i})+\sum_{k=1}^{n} \alpha_{k}(t_{j}) C_{k}(E_{i}),
\label{eq:pca}
\end{equation}
where  $\bar{F}(E_{i})$ is the time averaged flux at energy $E_{i}$
 (the time averaged spectrum is used as the origin for the eigenvector coordinate system).

 Eq.~\ref{eq:pca} amounts to a linear decomposition into
  $n$ independent components of  the variability (the eigenvectors $C_{1}$ , 
 $C_{2}$,...,$C_{n}$).
The normalisation coefficients of each PCA component
 (respectively $\alpha_{1}$, $\alpha_{2}$,..., $\alpha_{n}$) vary in time.
 Their fluctuations account for the sample variance. 
 On the other hand the eigenvectors $C_{k}$  are constant, they define the variability mode 
 of each PCA component. Since the eigenvectors describe fluctuations around 
 the time-average flux, both the $\alpha_{k}$ 
 and $C_{k}$ coefficients can take negative as well as positive values.
  If  the $C_{k}$ coefficients are positive
  at all energies, the variability mode associated to component $k$
  can be understood as due to an additive spectral component with a fixed shape 
  and a variable normalisation. 
  When the sign of $C_{k}$ depends on energy, 
  this corresponds to more complex spectral variability modes (e.g. pivoting).}

 As mentioned above, the PCA components are ordered according to the amount 
 of sample variance they account for (i.e. the observed fluctuations of $\alpha_{1}$ 
 cause more variance than those of $\alpha_{2}$ 
  which produce more variance than $\alpha_{3}$ etc...). 
The first few Principal Components (those representing most of the
variance in the data) should reveal the shape of the relevant spectral
components or variability modes. The weaker Principal Components might be expected
to be dominated by the statistical and systematic noise in the spectra.
To summarise, PCA finds the decomposition that maximises the variability 
due to lower order components, so that most of the variability can be described using
a small number of components.

 \subsection{Results of the PCA analysis}
 
The results of our  PCA analysis of the spectral variability of \object{Cygnus~X-1}
 are illustrated in Fig.~\ref{fig:pcatot}, which shows 
how the 2 first principal components
 affect the flux and spectrum and  their respective contribution
  to the total observed variance as a function of energy.
As can be seen on this figure, the first principal component (PC 1) 
consists in a variability mode dominated 
by variations in the luminosity (normalisation) with little change in the
spectral shape.  For this reason, in the following, we will refer to PC 1 as the 'flaring mode'.
 This component accounts for 68 \% of the sample
variance. 
Let us now consider the small spectral fluctuations induced by PC 1.
An increase in luminosity is associated with spectral hardening in the
3--10 keV band,  and a slight softening at higher energies which is particularly 
marked in the 10--30 keV band.  
This spectral variability suggests a variable power-law
spectrum with fixed spectral index moving on top of a constant thermal
emission disc emission plus reflection component.
Despite these small spectral variations, PC1 correlates very well
 with the high-energy flux.
A least square fit shows that the 3-200 keV flux relates to 
$\alpha_{1}$ through:
 \begin{equation}
 F_{3-200}=\bar{F}_{3-200}(1+7.64\alpha_{1}),
 \end{equation}
with a linear correlation coefficient of 0.98.
So that $\alpha_{1}$ can be viewed as a tracer of the hard X-ray luminosity of the source.

As shown in Fig~\ref{fig:pcatot}, the second PCA component (PC 2) 
can be described roughly  as a pivoting of the spectrum around 10 keV. 
The two spectra obtained for the minimum  and maximum values of the $\alpha_{2}$
 parameter controlling the strength of PC 2 are reminiscent of 
 the canonical LHS and HSS spectra. This component is responsible for
27~\% of the sample variance, and will be referred as the 'pivoting mode'.
$\alpha_{2}$ can thus be seen as a tracer of the hardness of the high-energy spectrum.
 
 The third PCA component (PC 3) 
 accounts for only 2~\% of the sample variance.
It  is dominated by fluctuations of the  relative normalisation of the spectrum in the 
  {\it JEM-X} band and {\it ISGRI}/{\it SPI} energy range.
   This component is most likely an instrumental artefact due to known calibration
    issues related to the dithering observation mode.
Indeed, since the pointed direction changes between successive 
 science windows and as the effects of vignetting and background non-uniformity
  are not perfectly corrected in the present release of the data analysis software, 
  this results in spurious variability that affects mainly the relative flux normalisation 
  between instruments. 
  This effect is particularly strong in {\it JEM-X} for which systematic errors 
  as large as 30 \% are expected for a source 5 degrees off axis.
  PCA enables us to disentangle such instrumental effects from the intrinsic variability. 
  Because they are independent of the intrinsic source variations 
  and as long as they are not dominant, they are filtered and isolated 
  into higher order PCA components.
  
  Similarly variability due to statistical noise is filtered out into higher order components.
   From the statistical uncertainties 
  on the measured count rates we expect 3 \%  of the observed variance to be due to statistical
   noise. 
  As the first 3 PCA components already account for 97 \% of the sample variance, the
   higher order components are most likely due to noise. 
   Fig.~\ref{fig:pcarms} shows the contribution of PC 1, PC 2, PC 3 
   and statistical noise to the observed variance spectrum  
   (i.e. the measured variance as a function of photon energy).
 These 4 components are enough to account for the observed variance within
  a few percents at all energies.

Since PC 3 is likely to be an instrumental artefact and  higher order components
 are probably noise, 
we conclude that the intrinsic source variability is largely dominated by PC 1 
(flaring mode) and PC 2 
(pivoting mode). Fig.~\ref{fig:pcavar} shows the time
evolution of the PCA parameters $\alpha_{1}$ and  $\alpha_{2}$.
$\alpha_{1}$, which traces the changes in bolometric luminosity at
nearly constant spectra shows important variability on time scales of order of a
few hours or less, but no clear systematic trend during the 4 days of
observation. In contrast, $\alpha_{2}$, which roughly traces the hardness
of the spectrum, seems to vary on longer time scales: it jumps  during
the first 2  days then decreases
in the second part of the observation.   This suggests that the physical
mechanisms responsible for PC 1 and PC 2 are distinct 
(which is also expected from the fact that, by construction,
 PC 1 and PC 2 are linearly independent) and apparently acting on
different time scales.

\begin{figure}[!t]
\centering
\includegraphics[width=\linewidth]{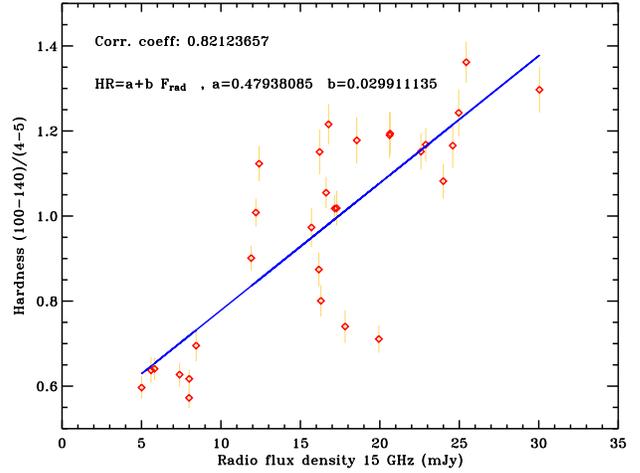}
\caption{Correlation between radio flux density and (100-140
keV)/(4-5 keV) hardness \label{fig:radiohr}}
\end{figure}

\begin{figure}[!ht]
\centering
\includegraphics[width=\linewidth]{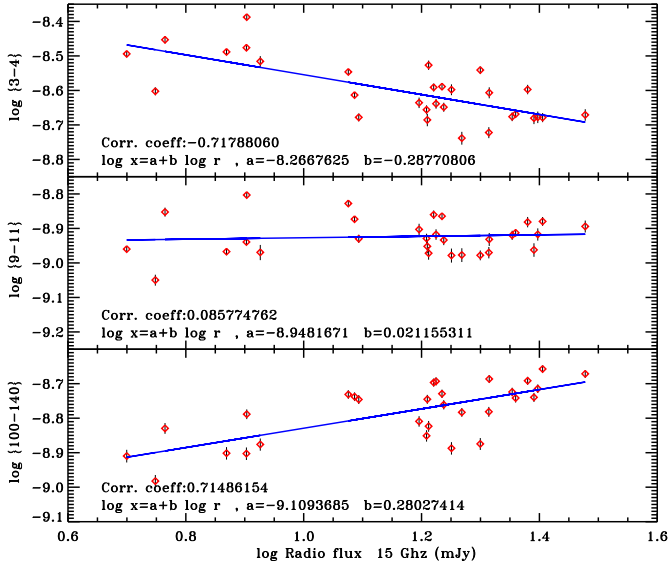}
\caption{Flux (ergs cm$^{-2}$ s$^{-1}$) in different energy bands (as indicated in keV),
versus 15 GHz radio flux density \label{fig:radiocor}}
\end{figure}

\subsection{Radio/high-energy correlation}

In order to study the possible correlations between the radio and hard X-ray emission,
we selected the science windows for which we had
 simultaneous radio pointings. When there were several radio pointings falling within
  a given science window we use the mean radio flux.
  The resulting light curve is shown on the bottom panel of Fig.~\ref{fig:pcavar}.
  A comparison with the time evolution of $\alpha_{1}$ and $\alpha_{2}$ indicates
   that the radio flux tends to follow the evolution of the pivoting mode.
    In other words, the radio flux tends to be stronger when the hard X-ray spectrum
     is harder.  This is clearly seen on Fig.~\ref{fig:radal} which shows that the 
     radio emission is strongly correlated to $\alpha_{2}$. The correlation is highly 
   significant. The Spearman rank test correlation coefficient is 0.78 corresponding
    to a probability that the correlation is by chance of $2 \times 10^{-7}$. 
    On the other hand, there is no hint of a correlation with the flaring mode as can
     be seen in the left panel of Fig.~\ref{fig:radal}. 
        
    It is also worth noting that the correlation between the pivoting mode and the radio flux 
    is valid only when we consider the data over more than one day, we find no evidence 
    for a correlation  on shorter time-scales.
     Therefore, our PCA analysis shows that on time scales of  days, the  radio jet
activity is correlated with hardness of the high-energy spectrum  
rather  than hard X-ray luminosity.

Obviously this can be seen directly from the light curves, although less comprehensively.
For instance Fig.~\ref{fig:radiohr} shows a strong correlation between 
the (100-140 keV)/(4-5 keV) hardness and the radio flux.
Fig.~\ref{fig:radiocor} shows  that the
 radio flux tends to be anti-correlated with the X-ray flux (3--7 keV)
keV)  and correlated with the soft gamma-rays ($>$15 keV). This
dependence of the {\it INTEGRAL}/radio flux correlation 
confirms both the presence of a pivot point located around 10 keV 
and the correlation of the radio luminosity 
 with the pivoting of the spectrum. 
 However, without PCA it would have been very difficult to demonstrate
 the presence of the 2 independent variability modes.
 PCA also brings a much clearer information on the spectral variability due to each mode, 
 and, on top of that, enables one to filter out systematic and statistical errors.

\begin{figure*}[!t]
\centering
\includegraphics[width=\linewidth]{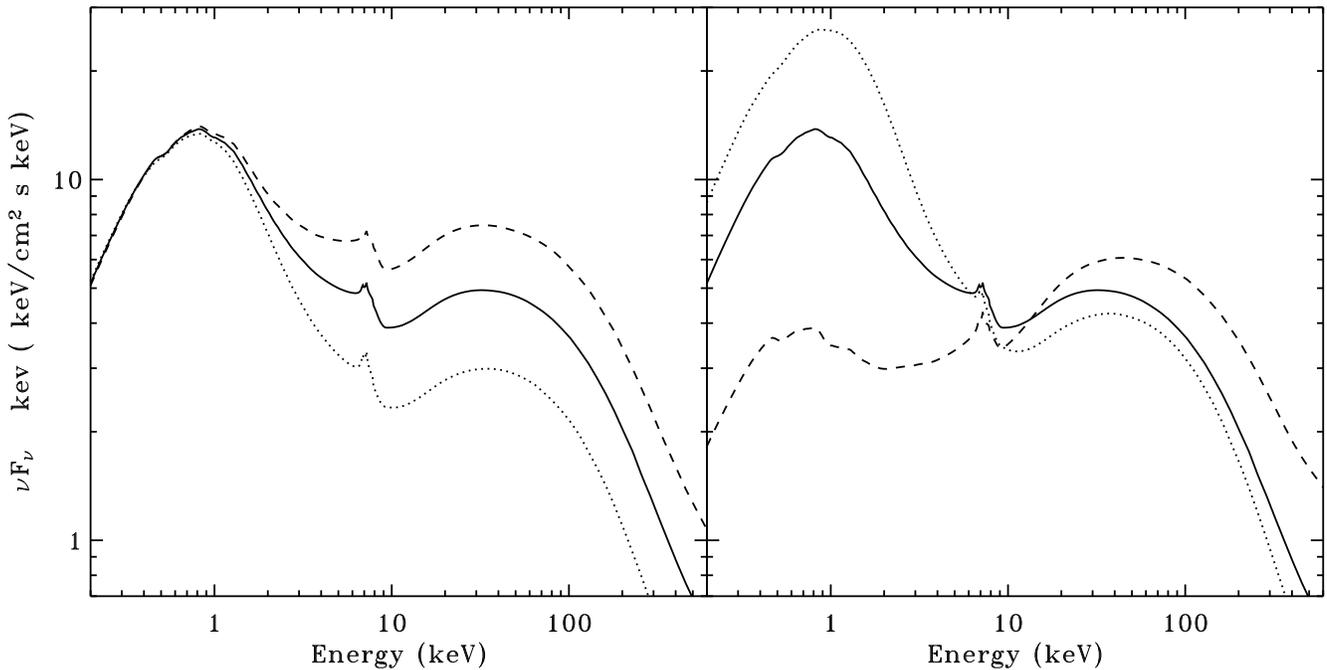}
\caption{Left panel: effect of varying $l_{h}$ by a factor of 2 on the {\sc eqpair} model 
with monoenergetic injection (see Sect.~\ref{sec:jointspec}). 
Solid curve: unabsorbed best-fit model ($l_{h}=8.5$); Dotted curve  $l_{h}=5.7$; 
 Dashed curve: $l_{h}=11.9$. 
 Right panel: effect of varying the soft photons flux by a factor of 8. 
 Solid curve: unabsorbed best-fit model  
 ($T_{\rm disc}=0.3$ keV; $l_{\rm h}/l_{\rm s}=0.85$). 
 Dotted curve: $T_{\rm disc}=0.357$ keV and $l_{\rm h}/l_{\rm s}=0.42$.
 Dashed curve: $T_{\rm disc}=0.212$ keV and $l_{\rm h}/l_{\rm s}=3.4$.}
 \label{fig:interp} 
\end{figure*}

\section{Discussion}

We have shown that, during our IMS observation, the variability of \object{Cygnus~X-1} 
can be described by two independent variability modes: 
\begin{itemize}
\item On time-scales of a few hours or less there are important changes
 in luminosity with little spectral variations (flaring mode).
 \item On longer time scales there is a spectral evolution with the 
 the spectrum pivoting around 10 keV.
 \end{itemize}
 We further showed that while there is no hint for a correlation between 
 the radio flux and the flaring mode,
 the radio is strongly correlated with the pivoting of the spectrum,
  in the sense that the radio flux
  is stronger when the hard X-ray spectrum is harder.
  This result strongly differs from what is usually reported in the LHS.
   Indeed, the radio flux is then
 positively correlated with the soft X-ray emission (3 -- 25 keV,
Corbel et al.  2000,
2003; Gallo, Fender, Pooley 2003). 
 
However, our results are not in conflict with these observations since
\object{Cygnus~X-1} was not in a typical LHS but in an IMS,
or rather was switching between different IMSs. 
Indeed, compilations of LHS and HSS spectra
   suggest that the spectral transition between LHS and HSS occurs
    through a pivoting around 10 keV (see e.g. Fig. 9 of McConnell et al 2002).
The evolution of the $\alpha_{2}$ parameter shown in Fig.~\ref{fig:pcavar} 
indicates that the source, initially in a 'soft' IMS, switched to a harder state 
during the first 2 days of observation and then transited back toward a 'soft' state.

Actually, the transition from LHS to HSS is known to be 
associated with a quenching of the radio emission (Corbel et al. 2000;
Gallo, Fender \& Pooley 2003). As the transition to the HSS
also corresponds to a strong softening of the spectrum, this is
consistent with the correlation between hardness and radio flux: 
when, during the observation, the source gets closer to the HSS 
the spectrum softens and simultaneously the radio flux decreases.
We also note that a recent analysis of Ryle and {\it RXTE} data of \object{Cygnus~X-1} (Gleissner et
 al., 2004) interestingly shows the same
correlation tendencies during failed state transitions (Ryle/ {\it RXTE/PCA}:
moderate anti-correlation, Ryle/ {\it RXTE/HEXTE}: correlation) as reported
here, albeit on timescales from weeks to years.

 It is interesting to speculate on the cause of the two variability modes. 
 We tried to reproduce such variability modes by varying the parameters  of the
 hybrid thermal/non-thermal Comptonisation models considered 
 in Sect.~\ref{sec:jointspec}.
 We used the best-fit model obtained with {\sc eqpair} 
 and mono-energetic particle injection. We choose this model because it gives 
 a better fit of the average {\it INTEGRAL} spectrum than the power-law injection model, 
 still avoiding the complication of an additional component that cannot be computed 
 self consistently. Actually, similar results are obtained for all of the 3 models we used 
 to fit the broad band spectrum.
 
  As shown in  the left panel of Fig.~\ref{fig:interp} it is possible to produce
   variations in luminosity by a factor comparable to what is observed
 and little spectral changes in the {\it INTEGRAL} band  by varying the coronal 
 compactness $l_{\rm h}$ by a factor of 2.
 In this context the flaring mode would correspond to variations of the dissipation rate
  in the corona possibly due to magnetic reconnection.
  This variability mode seems to be a characteristic of the HSS (Zdziarski et al. 2002).
  As we show here, it also provides a major contribution
   to the variability of the IMS.

Regarding the pivoting mode, it can be
 produced by changes in the flux of soft cooling photons at constant
  dissipation in the hot phase. 
 We performed simulations assuming that the accretion disc
  radiates  like a blackbody i.e.
  its flux $F_{disc} \propto l_{\rm s} \propto T_{\rm max}^4$ and constant $l_{h}$.
   For an increase of the disc temperature by a factor of 1.7, the disc luminosity 
   grows by a factor of 8. As in this model, the disc flux also corresponds
    to the soft cooling photon input in the corona and the heating ($\propto l_{\rm h}$)
     is kept constant, this leads to a steepening of the spectrum
      with a pivot around 10 keV of similar amplitude as in PC 2 
  (see Fig.~\ref{fig:interp}). 
   For the 1996 HSS, G99 found a ratio $l_{h}/l_{s}\sim0.3$ while
   in the LHS $l_{h}/l_{s}$ ranges between 3.5 to 15 (Ibragimov et al. 2005).
   The range of $l_{h}/l_{s}$ (0.4--3.4) required to reproduce the observed amplitude
    of the pivoting mode matches almost exactly the intermediate range between the HSS 
    and the lower limit of the LHS.  The source initially in a (quasi) HSS evolved
   toward the LHS but as soon as it was reached, it went back toward the HSS.
      
     Since,  in the {\it INTEGRAL} band,  the constraints on disc thermal emission are loose
 we did not attempt to model the data with a varying inner disc radius which is, moreover,
  difficult to disentangle from fluctuations of the disc temperature.
  In the fitted models as well as the models shown in Fig.~\ref{fig:interp}, 
  the inner disc radius is fixed at 6 $R_{g}$. 
   Nonetheless our result would also be consistent
   with the disc  moving inward and outward of the hot phase during the state transitions.
   Indeed, when the inner disc radius is approaching  the black hole, its maximum 
   temperature and luminosity increases\footnote{Unless the mass accretion rate is reduced
    by a larger amount, which seems very unlikely, the evidence being rather 
    that the accretion rate is often (but not always) larger in  the soft than in the LHS.}  leading to a more efficient cooling 
   of the hot flow/corona.  The anti-correlation between radio flux 
   and disc luminosity would be due to the jet expanding when the cold accretion
    disc recedes and then shrinking in the second phase of the observation 
    when the disc moves back inward.
   It is interesting to note that the change in disc flux required (a factor of $\sim 8$) 
   to explain the spectral evolution  
  is comparable to the amplitude of the variations of the radio flux
   (a factor of $\sim 6$). This suggests a direct relation between the disc flux and jet power.
   The overall change in bolometric luminosity occuring during the PC2 transition
    estimated from the  fiducial 'hard' and 'soft' state models shown on the left panel of Fig.~\ref{fig:interp}, 
    is about a factor of 2. Because of the relatively short time scale ($\sim$ a day) on which 
    the variation in luminosity occurs, it is unlikely to be driven by
     changes in the mass accretion rate. Most probably, it is  due to a change 
     in the radiative efficiency of the flow. The accretion flow could be less
     efficient in the LHS,  because about half of the accretion power
       is either swallowed by the black hole or pumped into the jet, while, in 
        the HSS, the cold disc is expected to be radiatively efficient.
      
      The contribution of the jet power to the total energetic
       output of black hole candidates in the LHS is a matter of debate.
        It has been argued that the jet could be strong and even dominant 
       (Fender et al. 2003;
        Malzac, Merloni \& Fabian 2004).
        The jump in luminosity by a factor of 2 that we infer during our mini state transition 
        sets an upper limit to the jet power, namely, 
        the jet power at a luminosity just below the transition is 
        at most comparable to the X-ray luminosity. 
        {  This uppper limit is in aggreement with a recent study of the jet interaction 
        with the surrounding interstellar medium by Gallo et al. (2005).
        According to these authors,  the jet power in \object{Cygnus~X-1} represents 
        between 6 to 100 \%  of the X-ray luminosity at the peak of the LHS.}
        In our model, the upper limit is reached if there is no advection into the black hole.
        In this case the mini-transition pivoting mode would correspond 
        to a redistribution of the accretion power 
        between the jet and the cold accretion disc.
        In this case all LHS sources would be jet dominated, since,
         according to the 
        jet and X-ray power scaling laws of  Fender et al. (2003), 
        the jet share of the energy budget is
         increasingly larger at lower luminosities.
         However, the amplitude of the luminosity changes we infer during the mini-transition (i.e. the factor of 2)
          is model dependent and also depends on the strength of the thermal disc contribution to the time-averaged spectrum, 
       which is poorly constrained with the INTEGRAL data. Similar observations with and even 
       broader spectral coverage (including soft X-rays) are required to consolidate this result.
         
         Moreover, although our data point toward an inefficient accretion flow in the LHS,
         it does not tell us about the cause of this inefficiency: pure advection as well 
         as  jet dominated accretion flows are both viable possibilities. 
          We note that a similar conclusion was reached by Chaty et al. 2003, on the basis of 
          the analysis of the multi-wavelength spectrum of the transient black hole binary 
          XTE J1118+480.

          The evolution of the hard X-ray corona luminosity during our IMS 
          observation is very puzzling. Indeed,  if, as commonly believed for the LHS, 
          the corona constitutes the base of the jet, it is difficult
            to conceive that changes in the jet power and/or extension is not associated
             to changes in the energetics of the corona.  The apparent lack of response of
             the radio jet to 
             the (short time-scale) X-ray fluctuations could well be due to the time delays
             required to propagate the information from the corona to 
             the distant radio emitting region, 
             which moreover have very different sizes. 
           However, one would still expect the corona/hot accretion flow 
           to  track the longer time-scale evolution of the radio jet and also 
           respond to changes in the disc power and/or distance of the truncation radius.           
           Instead, we infer dramatic changes in the jet and disc power that are anti-correlated
            with each other, but
     \emph{completely unrelated} to the fluctuations of the coronal power.
            On the one hand, the spectral pivoting described by PC 2 is understood 
            in terms of changes in the disc luminosity
           at \emph{constant} coronal luminosity. 
            And, on the other hand, the rapid fluctuations of the coronal 
            power that we do observe through PC 1 are apparently not associated 
            to fluctuations of the cold disc or jet emission. 
            The nature of the instabilities responsible for a coronal activity that is so uncoupled
             to the  jet and disc emission remains to be clarified. 
          { In any case, these results indicate that, in the IMS,
          the  corona does not play the same role as in the LHS,
         and the whole disc-corona-jet interaction  seem to work differently.}

\section{Conclusion}

The {\it INTEGRAL} IMS spectrum of \object{Cygnus~X-1} shows 
a high-energy cut-off or break around 100 keV. The shape of this cut-off 
differs from pure thermal Comptonisation, suggesting the presence of 
a non-thermal component at higher energies. 
The average broad band spectrum is well fitted 
with hybrid thermal/non-thermal Comptonisation models,
although some important parameters such as the fraction of non-thermal power
 are not well constrained because they 
 depend on the assumptions of the model (mono-energetic versus power-law injection, 
presence or absence of an additional soft component). Models with mono-energetic injection,
 or models with an additional soft component seem to be  favoured 
over standard power-law acceleration models.

 During our observation, 
the source presented a strong flux spectral variability occurring through 
2 independent variability modes:
\begin{itemize}
\item[i)] changes in the dissipation rate in the corona, due to local instabilities 
or flares, producing  a variability of the hard X-ray 
luminosity on time-scales of hours 
and no strong spectral alterations. 
Strikingly, this coronal activity seems to be unrelated to the evolution of the jet and cold disc luminosity.
\item[ii)] a slower 4-day evolution starting from a spectrum close to the canonical
 HSS toward an almost LHS and back. This spectral evolution was
  characterized by a pivoting of the spectrum around 10 keV. 
  It was correlated with the radio emission which was 
  stronger when the hard X-ray spectrum was harder.
  It is interpreted in terms of a variable soft cooling photon flux 
  in the corona associated with changes in the thermal disc luminosity and radio-jet power. 
  This interpretation suggests a jump in bolometric luminosity of about 
  a factor of 2 during the transition from LHS to HSS,
  which might indicates that the LHS accretion flow 
  is radiatively inefficient,
  half of the accretion power being possibly advected
   into the black hole and/or the radio jet.
\end{itemize}   
{ 
In the IMS, the jet power appears to be anti-correlated
    with the cold accretion disc luminosity, while the coronal power fluctuates 
    independently. Apparently, the coupling between accretion and ejection processes  
  differs from that of the LHS where the radio jet and the X-ray corona appear 
  intrinsically linked.   
 }
   
\begin{acknowledgements}
This paper is based  on observations with {\it INTEGRAL}, an ESA project with instruments and science data 
   centre funded by ESA member states (especially the PI countries: Denmark, France, Germany, Italy, Switzerland, Spain),
    Czech Republic and Poland, and with the participation of Russia and the USA. 
    The Ryle Telescope is supported by PPARC. JM acknowledges financial support from the
     MURST (COFIN98-02-15-41), PPARC, CNRS and European Union 
     (contract number ERBFMRX-CT98-0195, TMR network "Accretion onto black holes, 
      compact stars and protostars"). 
      This research was also supported in part by the National Science Foundation under 
      Grant No. PHY99-07949. 
\end{acknowledgements}

\end{document}